# Production of polyhydroxybutyrates and carbohydrates in wastewater-borne cyanobacteria: effect of nutrients limitation and photoperiods

**Dulce María Arias, Enrica Uggetti*, María Jesús García-Galán and Joan García**

GEMMA - Environmental Engineering and Microbiology Research Group, Department of Civil and Environmental Engineering, Universitat Politècnica de Catalunya-BarcelonaTech, c/ Jordi Girona 1-3, Building D1, E-08034, Barcelona, Spain

* Corresponding author:

Tel.: +34 934016465

Fax: +34 934017357

E-mail address: enrica.uggetti@upc.edu

**Abstract**

In this paper, different photoperiods and nutritional conditions are applied to a mixed wastewater-borne culture dominated by cyanobacteria in order to enhance polyhydroxybutirates (PHBs) and carbohydrates accumulation. To this end, two different experimental set-ups were performed. In the first set, the culture was submitted to permanent illuminance, whereas in the second the culture was submmited to light/dark alternation of 12h, testing in both cases two different nutritional regimes: a) N-limitation b) P-limitation. Results showed that the highest PHB concentration (104 mg $L^{-1}$) was achieved under P limited conditions and permanent illuminance, whereas the highest carbohydrates concentration (838 mg $L^{-1}$) was obtained under N limited condition and light/dark alternation. Regarding bioplastics and biofuel generation, this paper demonstrates that the accumulation of PHBs (bioplastics) and carbohydrates (potential biofuel substrate) is favoured in wastewater-borne cyanobacteria under nutrients limitation conditions.

1. **Introduction**

During the last decades, alternative energy sources such as biofuels, biogas and added-value products such as bioplastics have gained considerable attraction due to their potential to replace petroleum-based products and all their known drawbacks. Thus, the development of new sustainable and cost-effective technologies to obtain carbon neutral bio-products has nowadays become a priority [1]. In this context, special attention has been given to cyanobacteria due to their potential to synthesize a large variety of bioactive compounds and other valuable metabolites. Similarly to eukaryotic microalgae that accumulate starch, they can synthesize and store polysaccharides such as glycogen, but more interestingly, they also have the capacity to accumulate other compounds of interest such as polyhydroxybutyrates (PHBs) [2]. PHBs are polyesters synthesized as intracellular carbon and energy reserves. This family of polymers is characterized by their plastic-like chemical and physical properties, which added to their biodegradability and biocompatibility, makes them a promising alternative to petroleum-based plastics [3]. Currently, PHBs can be obtained by a number of different chemical and biotechnological processes, being fermentative routes from bacteria those more frequently used to produce and commercialize these polymers [4]. Nevertheless, these fermentative processes require a large amount of exogenous organic carbon addition and continuous oxygen supply, which nowadays makes this production of bacterial PHB much more expensive than the production of the conventional plastics [5].

On the other hand, glycogen is a water-soluble α-polyglycan which constitutes the primary and most suitable substrate for biofuel generation, mainly in anaerobic fermentation, anaerobic digestion and bio-hydrogen technologies [6,7]. The use of cyanobacteria to produce glycogen represents an advantage in comparison with other higher plants or green algae producing carbohydrates [8], due mainly to their lack of a hard cellulosic cell wall, which usually requires of additional pretreatment and further costly conversion processes to extract the product [9,10].

Most of the studies related to the production of PHBs and glycogen from cyanobacteria are based on pure or genetically modified cultures [6,11–14]. Such cultures involve the use of sterile medium substrates in expensive and highly controlled processes, which yield products still too expensive to compete with petroleum-based ones. In this context, a more sustainable alternative approach for the production of metabolites could be the use of mixed wastewater-borne cultures dominated by cyanobacteria. The possibility of maintaining a cyanobacteria dominated culture in a pilot photobioreactor fed with wastewater was recently demonstrated [15,16]. However, the metabolites production in this type of cultures is still limited and accumulation strategies should be further investigated.

Indeed, recent studies have demonstrated that intracellular concentration of either PHB or glycogen in cyanobacteria could be enhanced modifying environmental and cultivation factors such as temperature, pH, inorganic carbon availability, nutrients concentration (N and P) and light availability (photoperiod and intensity) [17]. Among these conditions, nutrients limitation is considered the most frequent approach for the enhancement of metabolites accumulation [7,11,12]. The lack of N and P has resulted in an increment of both PHB (up to 20% in terms of dry cell weight (dcw)) and glycogen content (up to 60% (dcw)) [7,18–21]. Other important factors to consider in metabolites production are the photoperiods and the light intensity, which affect crutial physiological processes such as photosynthesis, respiration, cell division and the intracellular carbon components [22,23].

The aim of this work is to apply different photoperiods and nutrients limitation conditions to a mixed wastewater-borne culture dominated by cyanobacteria in order to improve PHB and carbohydrates accumulation. The wastewater consortium was inoculated in synthetic growth medium in order to evaluate its PHB production under in N and P limiting mediums separatedly. To the author's knowledge, this is the first time that a cyanobacteria dominated mixed culture is enhanced to produce metabolites considering different photoperiods paired with nutrients limitation.

## 2. Materials and methods

*2.1 Reagents and chemicals*

$K_2HPO_4$, $NaNO_3$, $NaHCO_3$, $CaCl_2·2H_2O$, and $Na_2EDTA$ were obtained from Panreac (Barcelona, Spain), $MgSO_4·7H_2O$, $C_6H_8FeNO_7$, $C_6H_8O_7$, HCl, NaOH, chloroform ($CHCl_3$) and D-glucose were purchased from Scharlau (Barcelona, Spain). Methanol (MeOH), sulphuric acid ($H_2SO_4$), heptadecane ($C_{17}H_{36}$) and PHB-PHV co-polymer standard were purchased from Sigma-Aldrich (St. Louis, US). Glass microfiber filters (1 μm) were provided by Whatman (Maidstone, UK).

*2.2 Experimental set-up*

*2.2.1 Cyanobacteria dominated biomass*

The experimental set-up was located at the laboratory of the GEMMA research group (Universitat Politècnica de Catalunya. BarcelonaTech, Spain). Previously to this study, a consortium mostly formed by cyanobacteria (abundance 60-70%) cf. *Aphanocapsa* sp. and cf. *Chroococcidiopsis* sp. was selected and cultivated in a pilot-scale closed photobioreactor (PBR). The PBR (30L) was used as a tertiary wastewater treatment system fed with a secondary urban wastewater and liquid digestate, with a hydraulic retention time of 10 days. Detailed characteristics of this system can be found elsewhere [15]. The biomass dominated by cyanobacteria was collected from a harvesting tank connected to the PBR and it was thickened by gravity in laboratory Imhoff cones for 30 min before its use in this study.

*2.2.2 Experimental photobioreactors set-up*

Four batch tests were performed during two consecutive weeks (15 days) in order to improve both intracellular PHBs and glycogen production. They were carried out in four closed polymethacrylate cylindrical experimental PBRs with an inner diameter of 11cm, a total volume of 3 L and a working volume of 1 L (Fig. 1).

Experiments were carried out in two sets of two reactors each. In the first set, the effect of N and P limitation was tested under permanent illuminance; in the second set, the same nutrients conditions were tested under light/dark alternation (12h dark/12h light).

Right before the start of the experiments, 60 mL of settled biomass from the pilot-scale PBR was suspended within 1 L of growth medium in each of the four reactors. Microscopic images of the initial biomass are shown in Fig.2 and characteristics of the inoculum after suspension in the growth media are given in Table 1.

. In order to achieve N and P limitation, two different growth media were used:

- The two reactors with N limitation contained N-free BG-11 growth medium consisting of: 0.04 g $L^{-1}$ $K_2HPO_4$, 0.036 g $L^{-1}$ $CaCl_2·2H_2O$, 0.001 g $L^{-1}$ $Na_2EDTA$, 0.075 g $MgSO_4·7H_2O$, 0.01 g $C_6H_8FeNO_7$, 0.001 g $C_6H_8O_7$, and 1ml $L^{-1}$ of trace elements.

- The two reactors with P limitation contained P-free BG-11 growth medium consisting of: 1.5 g $NaNO_3$, 0.036 g $L^{-1}$ $CaCl_2·2H_2O$, 0.001 g $L^{-1}$ $Na_2EDTA$, 0.075 g $MgSO_4·7H_2O$, 0.01 g $C_6H_8FeNO_7$, 0.001 g $C_6H_8O_7$ and 1.0 ml $L^{-1}$ of trace elements.

Reactors were continuously agitated with a magnetic stirrer (Selecta, Spain) set at 250 rpm. Temperature was continuously measured by a probe inserted in the PBR (ABRA, Canada) and kept constant at 27 (±2) °C by means of a water jacket around the reactor. Continuous monitoring of pH was carried out with a pH sensor (HI1001, HANNA, USA) and kept at 8.7 with a pH controller (HI 8711, HANNA, USA) by the automated addition of HCl 0.1 N and NaOH 0.1 N. Light intensity was set at 220 μmol m$^{-2}$ s$^{-1}$. It was provided through two external halogen lamps (60 W) placed on two opposite sides of each PBR. Experiments were performed submitting cultures to two different nutrients limited conditions (N-limitation vs. P-limitation). In both experimental set-ups, NaHCO$_3$ was added manually to the cultures as the only inorganic carbon (IC) source, in order to provide enough carbon to be transformed into PHB/carbohydrate. Availability of NaHCO$_3$ was followed up by daily analyses of IC

*2.3 Analytical methods*

The cultures in the reactors were analyzed for total inorganic carbon (TIC), total organic carbon (TOC), orthophosphate (dissolved reactive phosphorus) (P-PO$_4^{3-}$), nitrite (N-NO$_2^-$) and nitrate (N-NO$_3^-$) on alternate days, 3 days per week. On the other hand, soluble inorganic carbon (IC) and soluble organic carbon (OC) were measured 5 days per week, analyzing samples previously filtered in a 1 μm pore glass microfiber filters. Total nitrogen (TN) and total phosphorus (TP) were measured 2 days per week. Dissolved organic nitrogen (DON) and dissolved organic phosphorus (DOP) were determined by the analysis of filtered samples following the same procedure used for TN and TP analysis respectively, and substracting the value of N-NO$_2^-$ and N-NO$_3^-$, in the case of DON, or, the value of P-PO$_4^{3-}$ in the case of DOP. Organic Nitrogen (ON) was calculated as the difference between TN and N-NO$_2^-$ and N-NO$_3^-$, whereas organic phosphorus (OP) was determined as the difference between TP and P-PO$_4^{3-}$. P-PO$_4^{3-}$, N-NO$^{2-}$, N-NO$^{3-}$ concentrations were analyzed using an ion chromatograph DIONEX ICS1000 (Thermo-scientific, USA), while TOC, TIC, OC, IC and TN were analyzed by using a C/N analyzer (21005, Analytikjena, Germany). TP was analyzed following the methodology described in 4500 B and 4500 P, respectively, of Standard Methods (APHA-AWWA-WPCF, 2001).

Total suspended solids (TSS) and volatile suspended solids (VSS) were measured in the culture three days per week following the gravimetric method 2540 C and 2540 D in Standard Methods (APHA-AWWA-WPCF, 2001). Chlorophyll *a* was measured two days per week using the procedure 10200 H described in the Standard Methods (APHA-AWWA-WPCF, 2001). Dissolved oxygen (DO) was measured daily with a dissolved oxygen-meter (Thermo-scientific, USA) respectively. DO was measured directly in each PBR, inserting the sensor in the mixed liquor.

*2.3.1. PHBs and carbohydrates analysis*

PHB and carbohydrates content were measured on a daily basis in the case of the constant illumination experiments, and at the end of both the light phase and the dark phase during the first week of the 12h light/dark experiments. 50 ml of mixed liquor was collected and centrifugated (4200 rpm, 10 min), posteriorly ultra-frozen at -80 ºC overnight in an ultra-freezer (Arctiko, Denmark) and finally freeze dried for 24h in a lyophilizer (-110 ºC, 0.049 hPa) (Scanvac, Denmark).

PHB extraction protocol was adapted from the methodology described by [24]. Briefly, approximately 2 mg of freezed-dried biomass were weighted in a glass tube with Teflon liner screw cap, where 1 mL of MeOH acidified with $H_2SO_4$ (20% v/v) and 1 mL of $CHCl_3$ containing 0.5 mg $mL^{-1}$ heptadecane were added as internal standards. The tubes were then incubated at 100 °C in a dry-heat thermo-block (Selecta, Spain) during 5 h. After this period, the tubes were cooled on ice for 30 min. Afterwards, 0.5 mL of deionized water was added and the tube was vortexed during 1 min to aid the two phase separation (MeOH and water in the upper phase and $CHCl_3$ in the lower phase). $CHCl_3$ that remained in the lower phase was then extracted with a Pasteur pipette and placed into a gas chromatography (GC) vial with molecular sieves to remove water. The co-polymer of PHB-PHV (86:14% wt, CAS 80181-

31-3) was used as a standard for HB and HV. A sixpoint calibration curve was prepared at different concentrations of PHB-PHV and processed in the same way as the real samples. PHB was determined by means of GC (7820A, Agilent Technologies, USA).

Carbohydrates content was extracted following the method described in [25]. Briefly, approximately 2 mg of freeze-driedbiomass were weighted and placed in glass tubes with Teflon liner screw caps, where 2mL of a diluted solution of 0.9 N HCl was added. The tubes were then incubated in a heating-block at 100 °C during 2h. The samples were cooled in an ice bath, the supernatant was extracted and then the carbohydrates content was determined following the phenol–sulfuric acid method described in [26], using *D*-glucose as a standard.

Since the biomass was initially composed by a mixed culture, composition changes within the reactors where examined by microscopy once a week for qualitative evaluation of microalgae populations. Microbial visualization was performed in an optic microscope (Motic, China) equipped with a camera (Fi2, Nikon, Japan) connected to a computer (software NIS-Element viewer®). Cyanobacteria and microalgae species were identified *in vivo* using conventional taxonomic books [27,28], as well as a database of cyanobacteria genus [29].

3. **Results and discussion**

*3.1 Biomass growth*

All the cultures remained in oxygenic conditions, thus, similar DO average values were found in the luminous phase during experiments in both photoperiods: 7.1±1.2 mg $L^{-1}$ and 7.6±1.9 mg $L^{-1}$ for N-limited and P-limited conditions respectively during permanent illuminance, and 7.3±0.5 mg $L^{-1}$ and 7.8±0.6 mg $L^{-1}$ for N-limited and P-limited conditions respectively during the light-dark alternation. Only a slight decrease was observed at the end of the dark phase (6.2±0.5 mg $L^{-1}$ and 6.6±0.5 mg $L^{-1}$ for N-limited and P-limited conditions, respectively).

Under constant illuminance, the biomass initial concentration of 0.35 g VSS $L^{-1}$ reached values up to 0.97 g VSS L$^{-1}$ on the 8$^{th}$ day of operation in the N-limited culture, which decreased to 0.86 g VSS L$^{-1}$ by the end of the experiment (Figure 3a). Results in the P-limited culture indicated a higher growth rate, achieving a concentration of 1.62 g VSS L$^{-1}$ also on the 8$^{th}$ day of operation, decreasing afterwards to 1.38 g VSS L$^{-1}$. On the other hand, under alternate illuminance, the initial biomass concentration of 0.33 g VSS L$^{-1}$ reached values up to 0.99 g VSS L$^{-1}$ on day 12, remaining stable till day 15 (0.76 gVSS L$^{-1}$) in the N-limited culture (Fig. 3b). Meanwhile, the P-limited culture showed an increasing trend during all the experimental time, achieving 1.35 g VSS L$^{-1}$ on day 15 of operation.

Chlorophyll *a* content in N-limited cultures showed a similar pattern under both illuminance conditions, having a clear decreasing pattern (Fig. 3c and 3d). Furthermore, these cultures developed a yellowish color during the experimental time, indicating the decay of pigments inside the reactors. In contrast, Chlorophyll *a* content in the P-limited culture under permanet illumination increased from 1.00 mg L$^{-1}$ (initial Chlorophyll *a* content) to a maximum of 2.61 mg L$^{-1}$ on day 10, following a very similar pattern to the biomass content. On the other hand, under alternate illuminance the initial concentration increased from 0.95 mg L$^{-1}$ until a maximum of 3.10 mg L$^{-1}$ on day 12. It is important to highlight that under P limitation in both photoperiods, Chlorophyll *a* content was higher under alternate illumination than in permanent illumination. This fact can be associated to the disruption of Chorophyll *a* biosynthesis caused by prolonged illumination periods as it was previously documented by [30,31].

Regarding biomass composition, under permanent illuminance in the experiment with N-limitation, the initial composition remained the same during all the experimental time. On the contrary, an increase in the amount of cf. *Aphanocapsa* sp. over the other species of green algae as well as other cyanobacteria genus was observed from the 8$^{th}$ day of operation onwards in the P-limitation experiment, as it can be seen in Fig, 4a. and Fig. S1 in Supplementary material. While, under alternate illuminance the evolution of the biomass composition in both N and P limitation experiments was very similar to that observed in the constant illuminance set-up. Hence, during N limitation, the same biomass composition throughout the time was observed, whereas in the culture submitted to P-limitation, an evident increase of cyanobacteria cf. *Aphanocapsa* sp. over the other species from day 8 onwards was registered (Fig. 4b).

These results demonstrate that, in both experimental set-ups, the initial biomass concentrations under both illuminance conditions increased 2.3-2.4 times fold with N limitation, and 4 times fold with P limitation, as a consequence of the light periodsand unlimited carbon supply. Furthermore, despite the evident increase of biomass, the observed growth was higher in both P limited cultures than in the N limited conditions.

*3.2 Nutrients' concentration*

As shown in Table 2, the initial TN concentrations ranged between 22 and 27 mg $L^{-1}$ in both N-limited experiments (constant illuminance and alternate, respectively) and between 339 and 344 mg $L^{-1}$ in the P-limited cultures. These values remained stable until the end of the experiments. The same was observed with the initial concentrations of TP, 14.7 mg $L^{-1}$ and 13.4 mg $L^{-1}$ in the N-limited cultures and 12.9 mg $L^{-1}$ and 10.6 mg $L^{-1}$ in the P-limited cultures. However, variations in the organic forms present in the reactors were observed according to the addition of IN or IP. In both set-ups under P-limitation, with no additional IP in the medium, the initial concentration of OP, averaging 11.8 mg $L^{-1}$, remained the same until the end of the experiment. A similar trend was observed in the N-limitation experiments, in which the final values remained close to the initial values (22.24 mg $L^{-1}$). On the contrary, when IN or IP were supplied to the cultures, an increase in ON and OP was observed in the corresponding reactors. Taken all that into consideration, it can be assumed that the nutrients supplemented to the culture were consumed and transformed into biomass. It is important to point out that the DON in this experiment ranged from 1.03 to 1.66 mg $L^{-1}$ in all the experiments, whereas DOP showed values below 0.67 mg L$^{-1}$ as it is referred for these types of systems [32]. All in all, all the ON and OP content mostly corresponded to the active biomass.

Regarding the OP/TOC ratio, it increased during the length of the experiments under all the nutritional and illuminance conditions. As shown in Table 3, the highest values were reached when the cultures were submitted to P limitation. Indeed, the ratio of OP with the biomass increased 3-5 times-fold. Under N-limitation, an increase around 2 times the initial value was achieved. The high differencebetween the initial and final ratios is attributed to the accumulation of intracellular carbon storage compounds.

Regarding the ON/TOC ratio, the intial value (3.8) increased only in the cultures under N limitation conditions in both set-ups. This results probably led to the reduction of Chlorophyll *a* content in these reactors observed from the third day of operation until the last day, as shown in Fig.3c and 3d. Furthermore, these cultures developed the yellowish color throughout the experimental time, an evident sign of chlorosis or bleaching, a process characterized by the degradation of pigments as Chlorophyll *a* [33]. Previous studies associated this process only to N limitation [34], although this phenomenon was also observed in the study of [35] in *Arthrospira* sp. submitted to P limited conditions. In this study, the different colors observed at a glance were confirmed in the microscopic images, showing an evident discoloration with respect to the initial culture when submitted to N limitation in both photoperiods (Fig. 4a and 4b).

*3.3 PHB production*

Under permanent illuminance, PHB concentration in the N-limited culture increased slowly till day 9, when it reached a value of approximately 50 mg L$^{-1}$. It remained constant until the end of the experiment (Fig. 5a); in contrast, PHB concentration in the P-limited culture reached a maximum of 104 mg L$^{-1}$ on day 8, which decreased to 90 mg L$^{-1}$ on day 9 and to 69 mg L$^{-1}$ on day 10. Concentrations oscillated between 60 and 80 mg L$^{-1}$ from that day until the end of the experiment.

Regarding the light/dark alternation experiments, concentration of PHB in the culture submitted to N-limited conditions slowly increased till day 12, achieving a concentration of 61 mg L$^{-1}$ which remained nearly constant till day 15 (Fig. 5b). A similar trend was observed in the the P-limited culture, reaching the same PHB concentration on day 12 that kept increasing till day 15 (76 mg L$^{-1}$).

Regarding the intracellular content, with permanent illumination the highest value observed under N limitation was 5.4% dcw and 5.7% dcw under P limitation. In contrast, maximum contents during the 12h light/dark periods were of 6.5% dcw under N limitation and 5.6% dcw under P limitation. In spite of this percentages, results indicate that highest concentrations of PHBs were reached when the culture was submitted to P limitation in both photoperiods, especially under permanent illuminance. Lower concentrations in the N-limitation experiments could be also influenced by the chlorosis observed in both photoperiod. This was also detected in the study of [36], who associated the delay of PHB accumulation in cyanobacteria *Spirulina platensis* to factors such as chlorosis influencing the pigment synthesis. In the present study, although the culture submitted to light/dark alternation and N limitation also presented chlorosis, similar and even higher PHB contents were reached (6.5% dcw) with respect to the P-limited conditions tested (5.6%-5.7% dcw) (Fig 5b, Table 4), if considering the intracellular content (%dcw) and not the concentration. This fact suggest that dark periods could improve PHB accumulation as previously found in *Nostoc muscorum* [37] and *Synechocystis* sp. PCC 6803 [38]. In these studies, the increase of PHB during dark periods was associated to the conversion of glycogen to PHB.

*3.4 Carbohydrates production*

Concerning carbohydrates, under constant illuminace the N-limited culture reached a maximum concentration of 641 mg L$^{-1}$ on day 8, whereas for the P-limited culture reached a maximum of 552 mg L$^{-1}$ on the same day. From that point onwards, concentrations decreased in both experiments, more markedly in the N-limitation set-up. (Fig. 6a). Under alternate illuminance, the N-limited culture accumulated a maximum concentration of 838 mg L$^{-1}$ on day 12, which rapidly decreased afterwards to 430 mg L$^{-1}$ at the end of the experiment, whereas the P-limited culture reached a maximum concentration of only 432 mg L$^{-1}$ on the 12$^{th}$ day of operation, which also decreased till day 15 (Fig. 6b).

Regarding intracellular contents, the highest carbohydrates accumulation was observed under N limitation, reaching a concentration of 63% dcw under permanent light, and 74% under 12h light/dark periods, in contrast to the maximum content of 46% and 35% dcw achieved with P limitation under permanent light and 12h light/dark respectively. As it is mentioned by [7], when cyanobacteria are submitted to N starvation, the flow of the photosynthetically fixed carbon is turned from the protein synthesis metabolic pathway to the lipid or carbohydrate synthesis pathways. In the present study, carbohydrates represented the major carbon storage form in the cultures compared to PHB. Indeed, the concentration as well as % dcw of carbohydrates reached are in the order of 8 times higher than those obtained of PHB. This higher accumulation of carbohydrates was also observed in the studies of [39,40].

*3.5 Metabolites production achievements*

The accumulation of carbohydrates and PHB in cyanobacteria submitted to nutrient limitation is a response to this stress condition. Thus, glycogen and PHB act as buffers to avoid useless metabolic cycles, especially during dark–light transitions, regulating the switch between photosynthetic and catabolic pathways in the cells [41]. In this study, biomass as well as metabolites concentrations reached the highest values under light/dark alternation with the only exception of PHB concentration under P limitation, which was higher under constant illuminance. The use of alternating cycles is more representative of natural conditions and it implies an advantage for further escalation of the process to outdoors systems, avoiding additional energy costs for illumination.

To our knowledge, this is the first study enhancing the accumulation of metabolites such as PHB and carbohydrates in mixed cyanobacteria-microalgae wastewater borne cultures. For comparison purposes, literature results on PHB and carbohydrates accumulation in photoautotrophic conditions on cultures of cyanobacteria are summarized in Table 4 and 5. As it can be observed, PHB and carbohydrates accumulation is species dependent and in the cases where the content of both polymers was evaluated (i.e. this study and the studies by [39,40], the accumulation of both seemed to follow different trends, as maximum concentrations were achieved after different incubation periods. In the case of PHB production, most of the studies reached values within a range of 0.2-8.5% dcw, which are near to the values found in this study. In some cases, as [37], a slightly higher accumulation percentage was obtained but after a longer time of incubation (21 days). In the study by [40], the strain *Synechocystis* sp. PCC6803 reached values above 13% PHB (dcw) after 12 days of experiments under both N and P starvation conditions, much higher than the maximum values of 5.4% and 5.7% obtained in this study under N and P limitation respectively and during the same incubation period. Only the studies performed by [42,43] obtained the highest percentages with cyanobacteria *Synechococcus* sp. MA19 submitted to nutrients limited conditions with inorganic carbon as the carbon source.

Regarding carbohydrates, it is important to highlight that they can be accumulated by both cyanobacteria and green algae. Thus, carbohydrates measured in this study included both glycogen accumulated by cyanobacteria and starch accumulated by green algae. Generally, maximum carbohydrates content was obtained under N-limitation in both photoperiods, with values of 63% dcw and 75% dcw under constant illuminance and light/dark alternation respectively. These results are similar and even higher than the maximum values found in other studies carried out under the same nutritional conditions and similar period of incubation, with the only exception of the study of [39], who obtained up to 70% dcw in 2.7 days. On the contrary, the carbohydrates content obtained in this work under P limitation (46% and 36% under constant illuminance and light/dark alternation, respectively) are higher than the values obtained in the studies of [39] (23% dcw after 2.7 d) and [40] (28.9% dcw after 12 d). Only the strain *Spirulina platensis* used by [13] and [44] was able to accumulate more than 60% dcw of carbohydrates in a P-limited culture.

In general, results obtained in this study reveal that cyanobacteria dominated cultures cultivated in wastewater effluents can be used as PHB and carbohydrates producers. The production of these valuable metabolites from wastewater native microorganisms could be a cost-effective alternative to pure cultures. Indeed, in this case the additional costs of biomass production and chemical inputs to maintain sterile conditions could be avoided if using waste streams as substrates. This fact involves a promising approach of biorefinery technology to produce either bioplastics or biofuels. The results showed herein highlight the need of further studies regarding the enhancement of the production of these by-products in these kind of cultures.

## 4. Conclusion

This study demonstrates the enhanced accumulation of both bioplastics (PHBs) and a potential biofuel substrates (carbohydrates), in a mixed wastewater-borne cyanobacteria dominated culture used for wastewater bioremediation. The effect of N and P limitation during two different photoperiods on metabolites production was evaluated during a period of two weeks. Results showed that the highest PHB concentration (104 mg $L^{-1}$) was reached under P limitation and constant illuminance, whereas the highest carbohydrates concentration off (838 mg L$^{-1}$) was obtained in N limitation under light/dark alternation. Regarding bioplastics and biofuel generation, this paper highlights and demonstrates that nutrients limitation could be a good approach to enhance PHB and carbohydrates accumulation in wastewater-borne cyanobacteria.

**Acknowledgments**

The authors would like to thank the European Commission [INCOVER, GA 689242] for their financial support. Dulce Arias kindly achnowledge her PhD scholarship funded by the National Council for Science and Technology (CONACYT) [328365]. M.J. García and E. Uggetti would like to thank the Spanish Ministry of Industry and Economy for their research grants [IJCI-2014-22767 and IJCI-2014-21594, respectively]. Authors appreciate Estel Rueda and Laura Torres for their contribution during the experiment deployment.

**References**

[1]  Lau N, Matsui M, Abdullah AA. Cyanobacteria: photoautotrophic microbial factories for the sustainable synthesis of industrial products. Biomed Res Int 2015;2015:1–9.

[2]  Stal L. Poly(hydroxyalkanoate) in cyanobacteria: an overview. FEMS Microbiol Lett 1992;103:169–80.

[3]  Reis MAM, Serafim LS, Lemos PC, Ramos a. M, Aguiar FR, Van Loosdrecht MCM. Production of polyhydroxyalkanoates by mixed microbial cultures. Bioprocess Biosyst Eng 2003;25:377–85.

[4]  Steinbüchel A, Füchtenbusch B. Bacterial and other biological systems for polyester production. Trends Biotechnol 1998;16:419–27.

[5]  Panda B, Jain P, Sharma L, Mallick N. Optimization of cultural and nutritional conditions for accumulation of poly-β-hydroxybutyrate in Synechocystis sp. PCC

6803. Bioresour Technol 2006;97:1296–301.

[6]  Aikawa S, Ho S, Nakanishi A, Chang J, Hasunuma T, Kondo A. Improving polyglucan production in cyanobacteria and microalgae via cultivation design and metabolic engineering 2015:886–98.

[7]  Markou G, Angelidaki I, Georgakakis D. Microalgal carbohydrates: An overview of the factors influencing carbohydrates production, and of main bioconversion technologies for production of biofuels. Appl Microbiol Biotechnol 2012;96:631–45.

[8]  Nozzi NE, Oliver JWK, Atsumi S. Cyanobacteria as a Platform for Biofuel Production. Front Bioeng Biotechnol 2013;1:1–6.

[9]  Bohutskyi P, Bouwer E. Biogas production from algae and cyanobacteria through anaerobic digestion: a review, analysis, and research needs. Adv. Biofuels Bioprod., New York: Springer; 2013, p. 873–975.

[10]  Mendez L, Mahdy A, Ballesteros M, González-Fernández C. Chlorella vulgaris vs cyanobacterial biomasses: Comparison in terms of biomass productivity and biogas yield. Energy Convers Manag 2015;92:137–42.

[11]  Drosg B. Photo-autotrophic Production of Poly(hydroxyalkanoates) in Cyanobacteria. Chem Biochem Eng Q 2015;29:145–56.

[12]  Koller M, Marsalek L. Cyanobacterial Polyhydroxyalkanoate Production: Status Quo and Quo Vadis? Curr Biotechnol 2015;04:1–1.

[13]  Markou G. Alteration of the biomass composition of Arthrospira (Spirulina) platensis under various amounts of limited phosphorus. Bioresour Technol 2012;116:533–5.

[14]  Meixner K, Fritz I, Daffert C, Markl K, Fuchs W, Drosg B. Processing

recommendations for using low-solids digestate as nutrient solution for poly-B-hydroxybutyrate production with Synechocystis salina. J Biotechnol 2016;240:61–7.

[15] Arias DM, Uggetti E, García-Galán MJ, García J. Cultivation and selection of cyanobacteria in a closed photobioreactor used for secondary effluent and digestate treatment. Sci Total Environ 2017;587-588:157–67.

[16] Van Den Hende S, Beelen V, Julien L, Lefoulon A, Vanhoucke T, Coolsaet C, et al. Technical potential of microalgal bacterial floc raceway ponds treating food-industry effluents while producing microalgal bacterial biomass : An outdoor pilot-scale study. Bioresour Technol 2016;218:969–79.

[17] Khajepour F, Hosseini SA, Ghorbani Nasrabadi R, Markou G. Effect of Light Intensity and Photoperiod on Growth and Biochemical Composition of a Local Isolate of Nostoc calcicola. Appl Biochem Biotechnol 2015;176:2279–89.

[18] Ansari S, Fatma T. Cyanobacterial polyhydroxybutyrate (PHB): Screening, optimization and characterization. PLoS One 2016;11:1–20.

[19] Kaewbai-ngam A, Incharoensakdi A, Monshupanee T. Increased accumulation of polyhydroxybutyrate in divergent cyanobacteria under nutrient-deprived photoautotrophy: An efficient conversion of solar energy and carbon dioxide to polyhydroxybutyrate by Calothrix scytonemicola TISTR 8095. Bioresour Technol 2016;212:342–7.

[20] Miyake M, Takase K, Narato M, Khatipov E, Schnackenberg J, Shirai M, et al. Polyhydroxybutyrate production from carbon dioxide by cyanobacteria. Appl Biochem Biotechnol 2000;84:991–1002.

[21] Panda B, Sharma L, Mallick N. Poly-β-hydroxybutyrate accumulation in Nostoc

muscorum and Spirulina platensis under phosphate limitation. J Plant Physiol 2005;162:1376–9.

[22] Krzemińska I, Pawlik-Skowrońska B, Trzcińska M, Tys J. Influence of photoperiods on the growth rate and biomass productivity of green microalgae. Bioprocess Biosyst Eng 2014;37:735–41.

[23] Renaud SM, Parry DL, Thinh L Van, Kuo C, Padovan A, Sammy N. Effect of light intensity on the proximate biochemical and fatty acid composition of Isochrysis sp. and Nannochloropsis oculata for use in tropical aquaculture. J Appl Phycol 1991;3:43–53.

[24] Lanham AB, Ricardo AR, Albuquerque MGE, Pardelha F, Carvalheira M, Coma M, et al. Determination of the extraction kinetics for the quantification of polyhydroxyalkanoate monomers in mixed microbial systems. Process Biochem 2013;48:1626–34.

[25] Lanham AB, Ricardo AR, Coma M, Fradinho J, Carvalheira M, Oehmen A, et al. Optimisation of glycogen quantification in mixed microbial cultures. Bioresour Technol 2012;118:518–25.

[26] DuBois M, Gilles KA, Hamilton JK, Rebers PA, Smith F. Colorimetric method for determination of sugars and related substances. Anal Chim Acta 1956;28:350–6.

[27] Bourrelly P. Les algues d'eau douce. Les algues vertes. 1st ed., Societé nouvelle des éditions doubée; 1985.

[28] Palmer CM. Algas en los abastecimientos de agua. Manual ilustrado acerca de la identificación, importancia y control de las algas en los abastecimientos de agua. México: Editorial Interamericana; 1962.

[29]  Komárek J, Hauer T. CyanoDB.cz - On-line database of cyanobacterial genera. - Word-wide electronic publication, Univ. of South Bohemia & Inst. of Botany AS CR. Http://www.cyanodb.cz/; 2013. [accessed, 17.03.14]

[30]  Sforza E, Simionato D, Giacometti GM, Bertucco A, Morosinotto T. Adjusted light and dark cycles can optimize photosynthetic efficiency in algae growing in photobioreactors. PLoS One 2012;7.

[31]  Li Z, Wakao S, Fischer BB, Niyogi KK. Sensing and Responding to Excess Light. Annu Rev Plant Biol 2009;60:239–60.

[32]  García J, Hernández-Mariné M, Mujeriego R. Analysis of key variables controlling phosphorus removal in high rate oxidation ponds provided with clarifiers. Water SA 2002;28:55–62.

[33]  Collier JL, Grossman AR. Chlorosis induced by nutrient deprivation in Synechococcus sp. strain PCC 7942: Not all bleaching is the same. J Bacteriol 1992;174:4718–26.

[34]  Sauer J, Schreiber U, Schmid R, Völker U, Forchhammer K. Nitrogen starvation-induced chlorosis in Synechococcus PCC 7942. Low-level photosynthesis as a mechanism of long-term survival. Plant Physiol 2001;126:233–43.

[35]  Markou G, Chatzipavlidis I, Georgakakis D. Carbohydrates Production and Bio-flocculation Characteristics in Cultures of Arthrospira (Spirulina) platensis: Improvements Through Phosphorus Limitation Process. Bioenergy Res 2012;5:915–25.

[36]  Jau MH, Yew SP, Toh PSY, Chong ASC, Chu WL, Phang SM, et al. Biosynthesis and mobilization of poly(3-hydroxybutyrate) [P(3HB)] by Spirulina platensis. Int J Biol Macromol 2005;36:144–51.

[37]  Sharma L, Mallick N. Accumulation of poly-β-hydroxybutyrate in Nostoc muscorum: Regulation by pH, light-dark cycles, N and P status and carbon sources. Bioresour Technol 2005;96:1304–10.

[38]  Wu GF, Wu QY, Shen ZY. Accumulation of poly-beta-hydroxybutyrate in cyanobacterium Synechocystis sp. PCC6803. Bioresour Technol 2001;76:85–90.

[39]  De Philippis R, Sili C, Vincenzini M. Glycogen and poly-β-hydroxybutyrate synthesis in Spirulina maxima. J Gen Microbiol 1992;138:1623–8. doi:10.1099/00221287-138-8-1623.

[40]  Monshupanee T, Incharoensakdi A. Enhanced accumulation of glycogen, lipids and polyhydroxybutyrate under optimal nutrients and light intensities in the cyanobacterium Synechocystis sp. PCC 6803. J Appl Microbiol 2014;116:830–8.

[41]  Gründel M, Scheunemann R, Lockau W, Zilliges Y. Impaired glycogen synthesis causes metabolic overflow reactions and affects stress responses in the cyanobacterium Synechocystis sp. PCC 6803. Microbiol (United Kingdom) 2012;158:3032–43.

[42]  Miyake M, Erata M, Asada Y. A thermophilic cyanobacterium, Synechococcus sp. MA19, capable of accumulating poly-B-hydroxybutyrate. J Ferment Bioeng 1996;82:512–4.

[43]  Nishioka M, Nakai K, Miyake M, Asada Y, Taya M. Production of the poly-beta-hydroxyalkanoate by thermophilic cyanobacterium, Synechococcus sp. MA19, under phosphate-limited condition. Biotechnol Lett 2001;23:1095–9.

[44]  Markou G, Angelidaki I, Nerantzis E, Georgakakis D. Bioethanol production by carbohydrate-enriched biomass of Arthrospira (Spirulina) platensis. Energies 2013;6:3937–50.

[45]  Lama L, Nicolaus B, Calandrelli V, Manca MC, Romano I, Gambacorta A. Effect of growth conditions on endo- and exopolymer biosynthesis in Anabaena cylindrica 10 C. Phytochemistry 1996;42:655–650.

[46]  Aikawa S, Izumi Y, Matsuda F, Hasunuma T, Chang JS, Kondo A. Synergistic enhancement of glycogen production in Arthrospira platensis by optimization of light intensity and nitrate supply. Bioresour Technol 2012;108:211–5.

[47]  Sassano CEN, Gioielli LA, Ferreira LS, Rodrigues MS, Sato S, Converti A, et al. Evaluation of the composition of continuously-cultivated Arthrospira (Spirulina) platensis using ammonium chloride as nitrogen source. Biomass and Bioenergy 2010;34:1732–8.

Table 1. Characterization of the inoculum (biomass) taken from the 30L PBR and added in the four experimental PBR containing growth medium (n=4). Values are given as mean values (standard deviation).

| Parameter | Mean value (Standard deviation) |
| --- | --- |
| Temperature (ºC) | 24 |
| pH | 8.2 (0.2) |
| TSS (g L$^{-1}$) | 0.43 (0.05) |
| VSS (g L$^{-1}$) | 0.34 (0.06) |
| Chlorophyll *a* (mg L$^{-1}$) | 1.00 (0.08) |
| TOC (mg L$^{-1}$) | 84.6 (7.5) |
| ON (mg L$^{-1}$) | 22.2 (1.8) |
| OP (mg L$^{-1}$) | 11.8 (1.6) |
| PHB (% (w/gTS)) | 3.4 (0.3) |
| Carbohydrates (% (w/gTS)) | 7.4 (0.4) |

Table 2. N and P values of the culture at the end of the experiment (day 15).

| Parameter [mg L$^{-1}$] | Permanent illuminance | | Light/dark alternation | |
|---|---|---|---|---|
| | N-limited | P-limited | N-limited | P-limited |
| TN | 22.28 | 339.22 | 27.22 | 344.05 |
| ON | 22.28 | 154.22 | 27.22 | 238.05 |
| IN | 0 | 185 | 0 | 106 |
| TP | 14.73 | 12.9 | 13.36 | 10.62 |
| OP | 12.13 | 12.9 | 11.66 | 10.62 |
| IP | 2.65 | 0 | 1.7 | 0 |

Table 3. ON/TOC and OP/TOC ratios in the different cultures registered at the end of the experiments (day 15).

| Parameter | Initial value* | Permanent illuminance | | Light/dark alternation | |
|---|---|---|---|---|---|
| | | N-limited | P-limited | N-limited | P-limited |
| ON/TOC | 3.8 (0.3) | 8.56 | 1.20 | 6.8 | 1.52 |
| OP/TOC | 7.1 (0.5) | 15.79 | 25.57 | 15.9 | 34.21 |

*Average values with standard deviation for the four reactors.

Table 4. Summary of the maximum percentages and concentration values of PHB in the experiments performed in this study compared with other cyanobacteria culture studies.

| Cyanobacteria cultivated | Nutrient limited | Photoperiod Light:dark (h) | Maximum concentration (mgL$^{-1}$) | Maximum (% dcw) | Day of incubation (d) | Reference |
|---|---|---|---|---|---|---|
| Cyanobacteria dominated mixed culture | N | 24:0 | 51.6 | 5.4 | 9 | This study |
| Cyanobacteria dominated mixed culture | P | 24:0 | 104.23 | 5.7 | 8 | This study |
| Cyanobacteria dominated mixed culture | N | 12:12 | 61.61 | 6.5 | 12 | This study |
| Cyanobacteria dominated mixed culture | P | 12:12 | 76.36 | 5.6 | 15 | This study |
| *Nostoc muscorum* | N | 14:10 | - | 6.4 | 21 | [18] |
| Anabaena cylindrica | N | 24:0 | - | 0.2 | 21 | [45] |
| *Synechococcus* sp. MA19 | N | 0:24 | 67.2 | 21 | 6 | [42] |
| *Synechocystis sp. PCC6803* | N | 24:0 | - | 14.6 | 12 | [40] |
| *Synechocystis sp. PCC6803* | P | 24:0 | - | 13.5 | 12 | [40] |
| *Synechocystis salina* | P | 16:8 | 123.2 | 6 | 30 | [14] |
| *Synechococcus* sp. MA19 | P | 24:0 | 1400 | 62 | 4 | [43] |
| *Spitulina maxima* | N | 24:0 | - | 0.7 | 4 | [39] |
| *Spitulina maxima* | P | 24:0 | - | 1.2 | 4 | [39] |
| *Nostoc muscorum* | P | 14:10 | - | 8.5 | 21 | [37] |
| *Spitulina platensis* | P | 14:10 |  | 3.5 | 60 | [5] |

Table 5. Summary of the maximum percentages and concentration values of carbohydrates in the experiments performed in this study compared with other cyanobacteria culture studies.

| Cyanobacteria cultivated | Nutrient limited | Photoperiod | Maximum concentration (mgL$^{-1}$) | Maximum (% dcw) | Days of incubation (d) | Reference |
| --- | --- | --- | --- | --- | --- | --- |
| Cyanobacteria dominated mixed culture | N | 24:0 | 641.30 | 62.71 | 8 | This study |
| Cyanobacteria dominated mixed culture | P | 24:0 | 662.38 | 46.05 | 12 | This study |
| Cyanobacteria dominated mixed culture | N | 12:12 | 838.05 | 74.76 | 12 | This study |
| Cyanobacteria dominated mixed culture | P | 12:12 | 432.13 | 35.98 | 12 | This study |
| *Arthrospira platensis* | N | 24:0 | 800 | 65 | 3.5 | [46] |
| *Spitulina platensis* | P | 24:0 | - | 65 | - | [35] |
| *Spitulina platensis* | P | 24:0 | - | 63 | 9 | [44] |
| *Synechocystis sp.* PCC 6803 | N | 24:0 | - | 36.8 | 12 | [40] |
| *Synechocystis sp.* PCC 6803 | P | 24:0 | - | 28.9 | 12 | [40] |
| *Spitulina maxima* | N | 24:0 | - | 70 | 2.7 | [39] |
| *Spitulina maxima* | P | 24:0 | - | 23 | 2.7 | [39] |
| *Arthrospira platensis* | N | 24:0 | - | 65 | 7 | [47] |

**List of figures**

Figure 1. Schematic diagram of each photobioreactor (PBR) set-up: a) Body of the PBR, b) cover, c) water jacket; arrows indicate the water flux by the water jacket around the PBR, d) external lamps, e) magnetic stirrer, f) pH sensor, g) pH controller, h) acid solution, i) basic solution, j, temperature sensor, k) tube for manual addition of carbon.

Figure 2. Microscopic images illustrating the initial microbial composition of the culture. a), b) mixed culture dominated by cyanobacteria immersed in flocs observed in phase contrast microscopy (200X) and (400X) respectively; note darker cyanobacteria aggregates; c), d) detail of floc composed by cyanobacteria *Aphanocapsa* sp. and *Chroococcidiopsis* sp. (bigger and darker cells than *Aphanocapsa* sp.), green algae *Chlorella* sp., and diatoms observed in bright light microscopy (1000X).

Figure 3. Biomass (VSS) and Chlorophyll a concentration under N and P limitation in the cultures submitted to a) and c) permanent illuminance and b) and d) light/dark alternation.

Figure 4. Microscopic images illustrating the microbial composition evolution of the culture submitted to a) permanent illuminance and b) alternate illuminance under N and P limitation trough the time. Microscopy technique used is indicated below each picture.

Figure 5. PHB concentration under N and P limitation in the cultures submitted to a) permanent illuminance and b) light/dark alternation.

Figure 6. Carbohydrates concentration under N and P limitation in the cultures submitted to a) permanent illuminance and b) light/dark alternation.

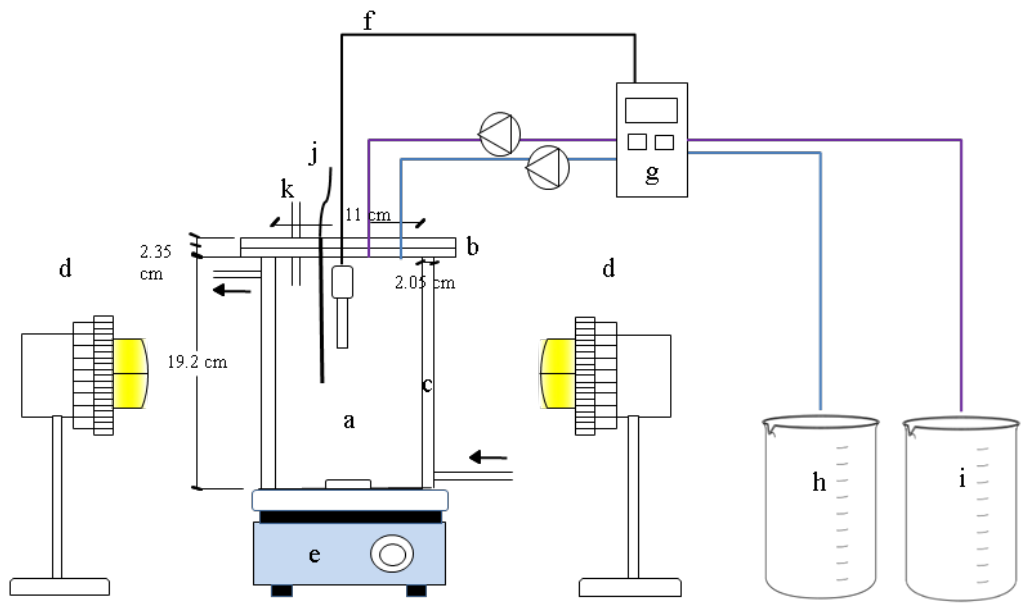

Fig. 1

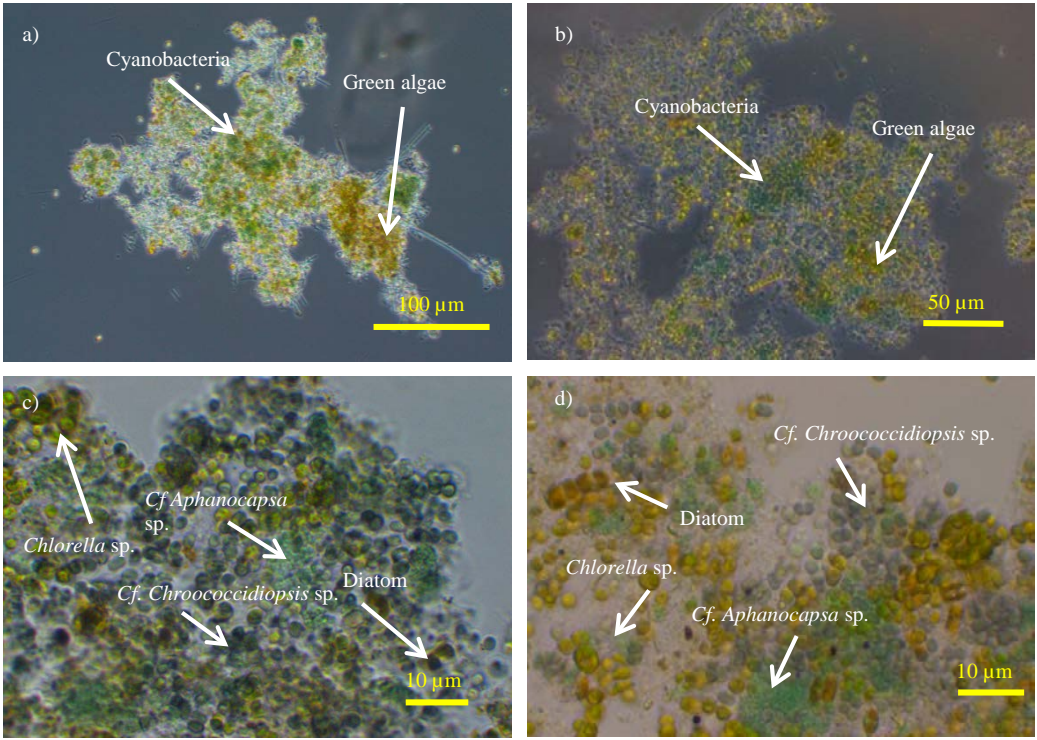

Fig. 2

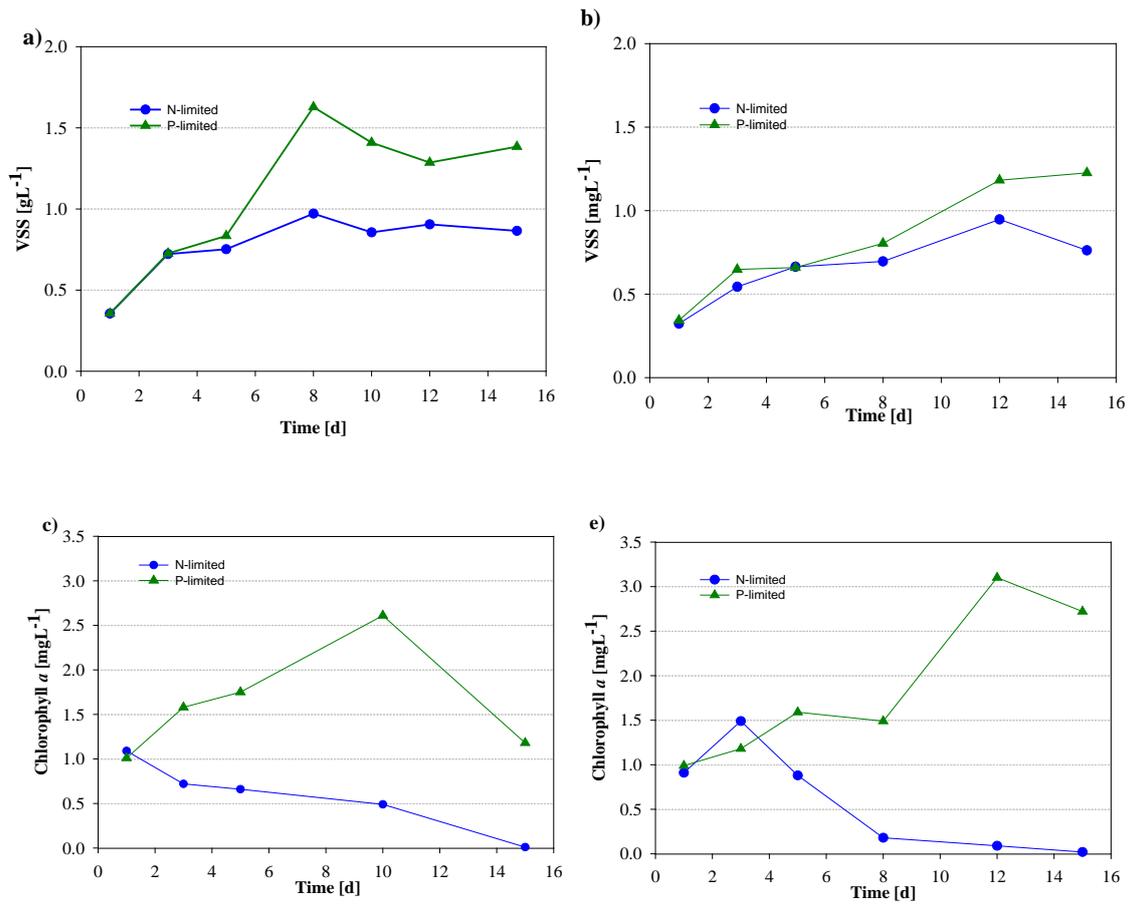

Fig. 3

| a) | | Day | | | |
|---|---|---|---|---|---|
| | | 3 | 8 | 12 | 15 |
| Constant illumination | N-limited | 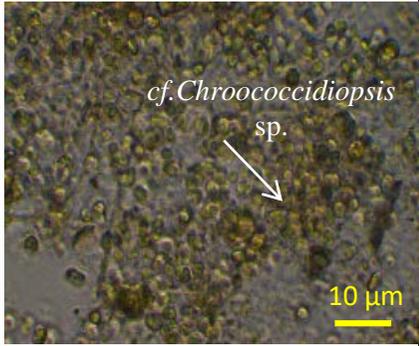 Bright light microscopy (1000x) | 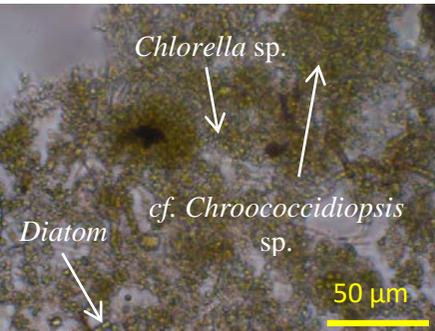 Bright light microscopy (...) | 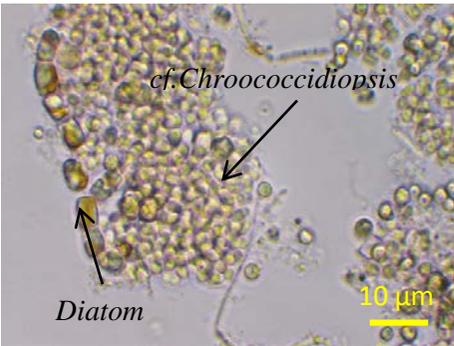 Bright light microscopy (1000x) | 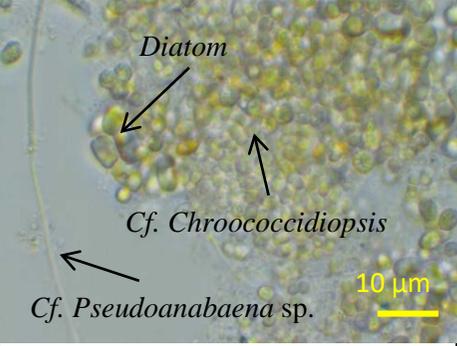 Bright light microscopy (1000x) |
| | P-limited | 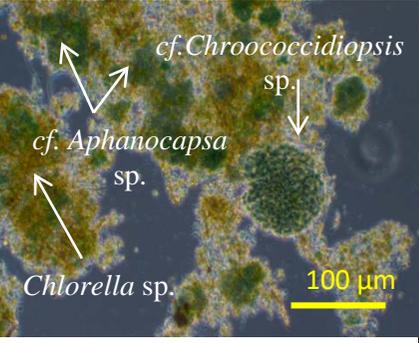 Phase contrast microscopy (200x) | 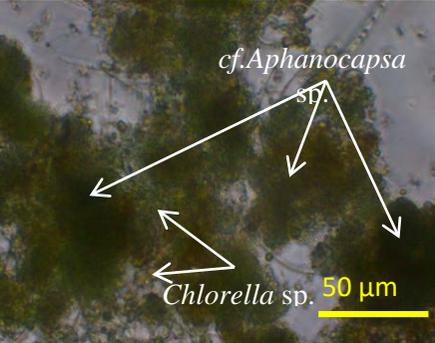 Bright light microscopy (400x) | 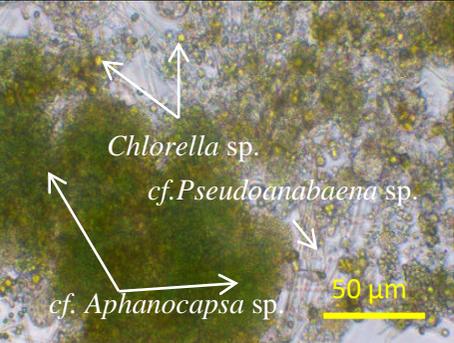 Bright light microscopy (400x) | 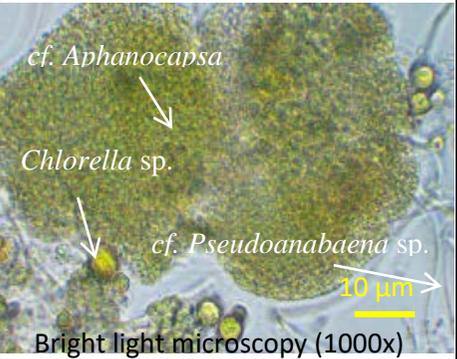 Bright light microscopy (1000x) |

| | | Day | | | |
|---|---|---|---|---|---|
| | | 3 | 8 | 12 | 15 |
| Alternate illumination | N-limited | Bright light microscopy (1000x) — *Chlorella* sp., *Diatoms*, *cf. Chroococcidiopsis* sp. | Bright light microscopy (400x) — *cf. Psudoanabaena* sp., *cf. Chroococcidiopsis* sp. | Bright light microscopy (1000x) — *cf. Chroococcidiopsis* sp., *cf. Psudoanabaena* sp. | Bright light microscopy (1000x) — *Chlorella* sp., *cf. Chroococcidiopsis* sp. |
| | P-limited | Phase contrast microscopy (200x) — *cf. Chroococcidiopsis* sp., *cf. Aphanocapsa* sp., *Chlorella* sp., *Diatoms* | Bright light microscopy (400x) — *cf. Aphanocapsa* sp., *Chlorella* sp. | Bright light microscopy (400x) — *Chlorella* sp., *cf. Aphanocapsa* sp. | Bright light microscopy (1000x) — *Cf. Aphanocapsa* sp., *Chlorella* sp. |

b)

Fig 4

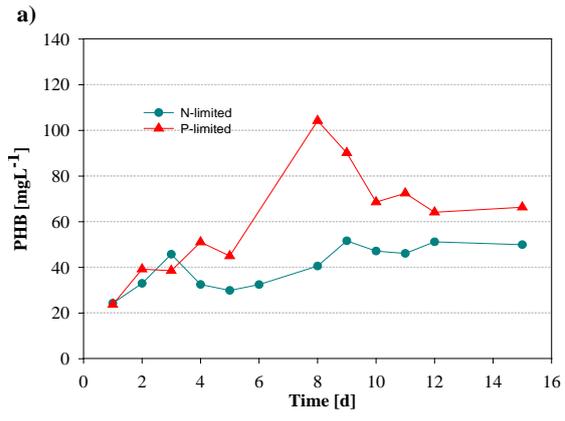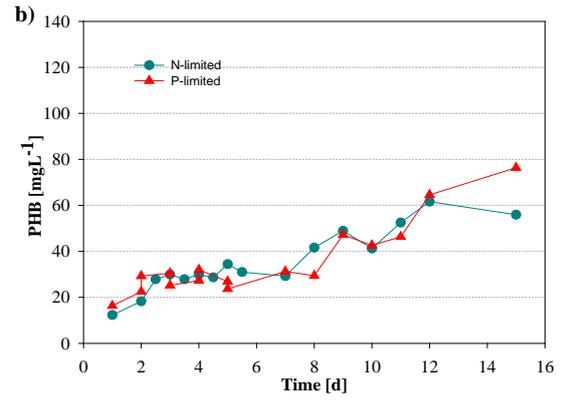

Fig. 5

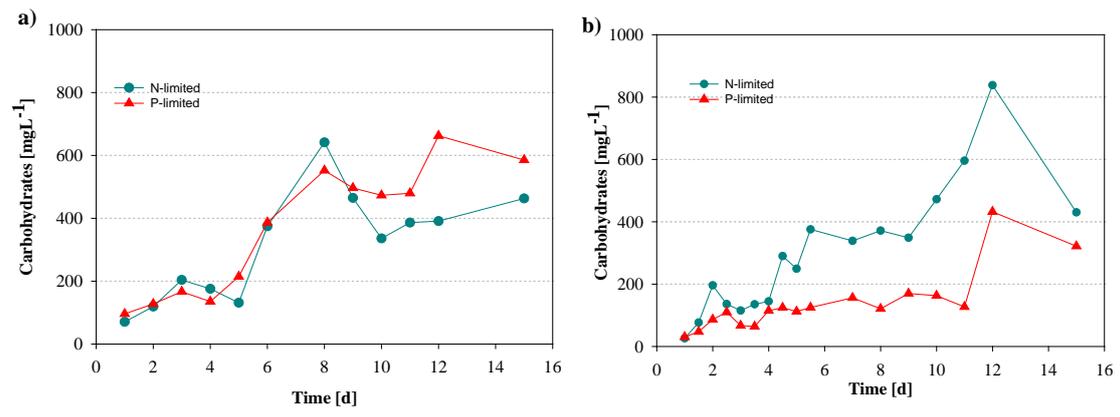

Fig. 6

**Supplementary material**

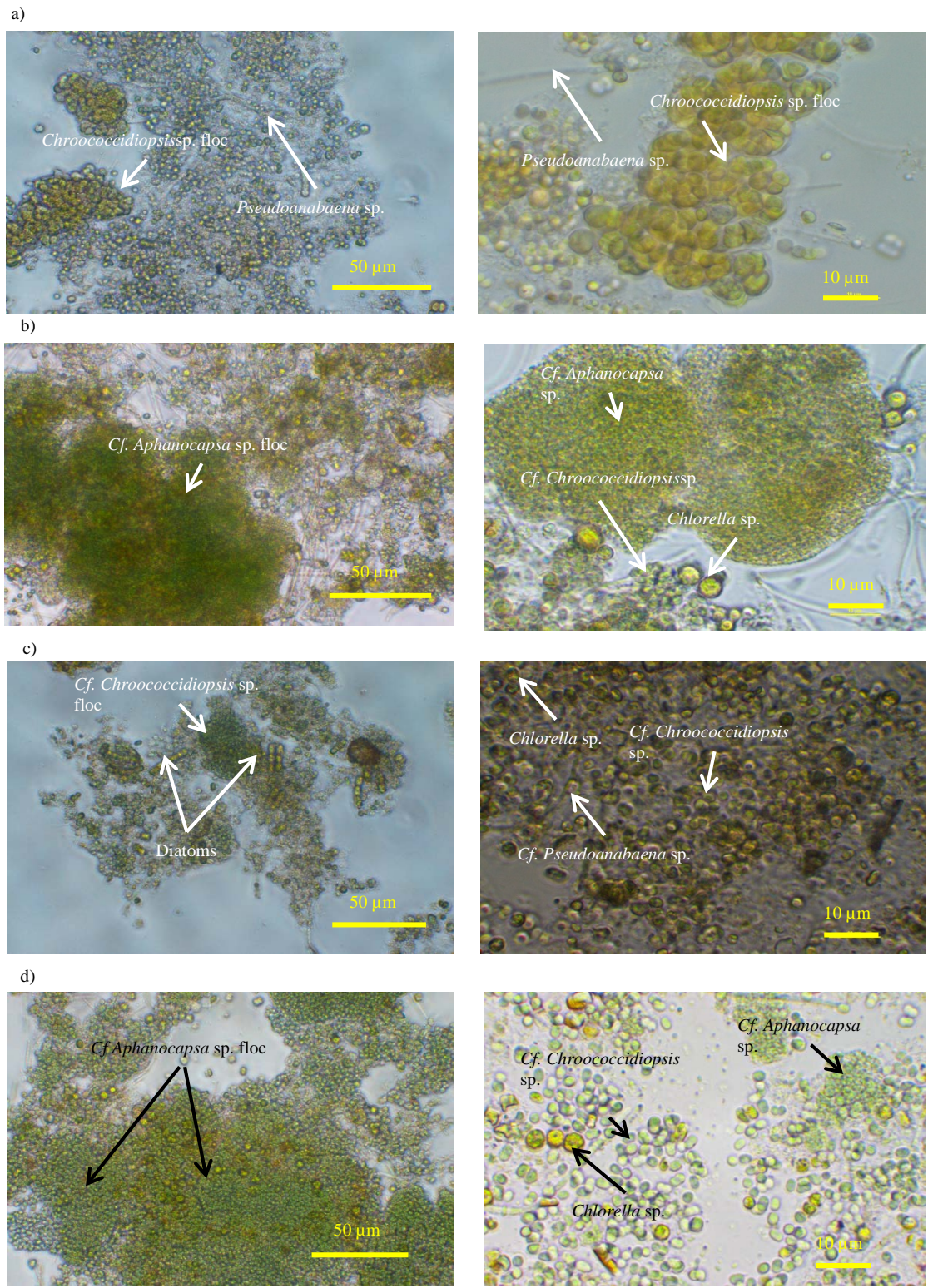

Fig. S1. Microscopic images illustrating the general view (left) at 400x and detailed view (right) at 100x of the microbial composition of the culture in the end of the experimental

time. a) N-limited culture under permanent illuminance showing cyanobacteria floc with cf. *Chroococcidiopsis* sp. and a filaments of cf. *Pseudoanabaena* sp.; b) Culture submitted to P limitation under permanent illuminance showing large colonies of cf. *Aphanocapsa* sp. immersed in flocs, with some filaments of cf. *pseudoanabaena* sp. and dispersed *Chlorella* sp.; c) N-limited culture under light/dark alternation showing cyanobacteria dominated floc composed by cf. *Chroococcidiopsis* sp. and some filaments of cf. *Pseudoanabaena* sp. and diatoms immersed; d) P-limitated culture under light/dark alternation showing large flocs composed by cf. *Aphanocapsa* sp. and cf. *Chroococcidiopsis* sp.